# Beam-based mapping of pulsed septum stray field at VEPP-2000 collider

D. Shwartz[1,2,*], I. Koop[1,2], M. Lyalin[2], E. Perevedentsev[1,2]

[1]*Budker Institute of Nuclear Physics, 630090 Novosibirsk, Russia*
[2]*Novosibirsk State University, 630090 Novosibirsk, Russia*

Various beam-based methods are widely used for machine study and accurate tuning. Beam closed orbit (CO) responses are routinely used worldwide for magnetic elements alignment and lattice correction. The paper presents the beam-based field mapping technique, applied for measurement of pulsed septum magnet stray field at VEPP-2000 electron-positron collider. Relatively slow pulse of weak leakage field disturbs the CO allowing detecting field longitudinal integral. While scanning with transverse coordinate in wide range one can measure the field map of the localized perturbation at machine in situ.

## I. INTRODUCTION

Pulsed septum magnets are used widely for beam injection/extraction [1-6] in order to provide high field, low power consumption and compactness. Two types of septa are recognized: direct-drive or eddy-current. The latter type can be considered as a quasi-coaxial cable that results in confinement of magnetic field inside the return conductor and low stray fields. Nevertheless, if we are interested in control of low (comparatively to guide septum field) and long-term (comparatively to driving pulse) leakage fields the problem becomes very difficult for any design of magnet.

At VEPP-2000 electron-positron collider [7] the eddy current type septa are used for beams injection. It was found that low but inhomogeneous stray field disturbs the circulating beam and affects top-up injection efficiency. To study the time and space field structure the beam-based method was proposed.

Various beam-based methods are used for storage ring studies. The beam closed orbit response for applied perturbation usually used for magnetic fields study along the reference closed orbit. Now we propose to scan with CO across the full aperture to measure the field map of the localized perturbation.

Similar studies were already carried out earlier at New-SUBARU storage ring but the corresponding report [8] is concentrated on the measured leakage field compensation while the measurement technique is not presented in details. In this paper we discuss the problems and precision of mapping leakage field distribution at the level of several Gauss that is less than $10^{-3}$ of septum guiding field.

## II. VEPP-2000 INJECTION

The schematic view of VEPP-2000 storage ring with transfer channels is presented in Fig. 1. Machine description and basic parameters can be found elsewhere [7,9-11]. It is a small, 24 m perimeter, single-ring electron-positron collider with energy range of 150-1000 MeV per beam. It operates with round beams and uses solenoids for final focusing [12].

The injection scheme [13] is dictated by very tight room. Although the additional powerful pulsed magnet (ME4/MP4) is introduced to relax the septum magnet ME5/MP5, the latter should produce field over 20 kGs at collider top energy of 1 GeV.

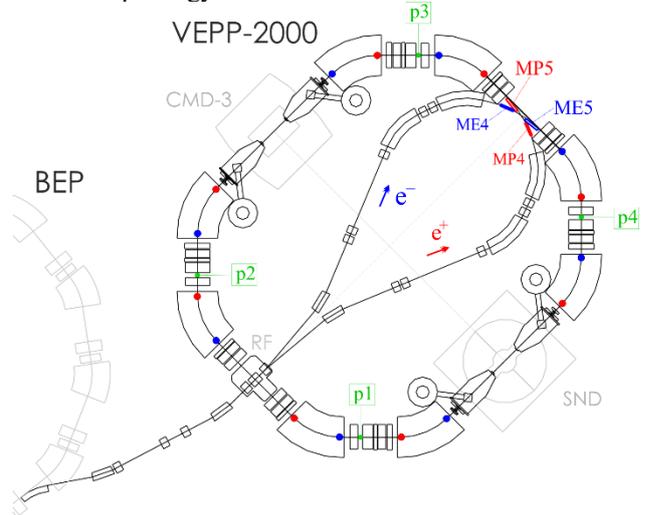

FIG. 1: VEPP-2000 collider overview. The electron injection septum ME5 and additional pulsed magnet ME4 are shown with blue, the positron injection elements shown with red. With green four pickups position is marked. Blue and red points along the beam orbit indicate the synchrotron radiation outputs for electron and positron beam imaging correspondingly.

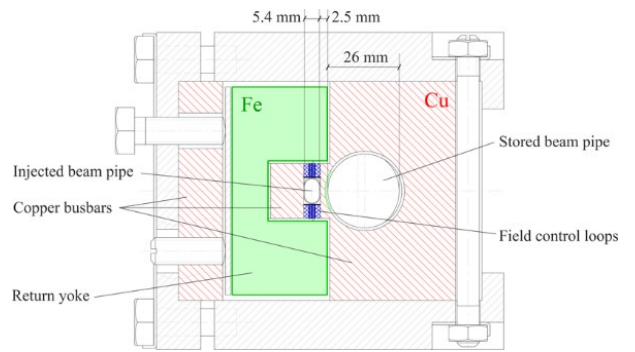

FIG. 2: Septum magnet cross-section.

The septum magnet is an eddy-current type magnet with full-sine driving current pulse. The pulse rising time is T/4 = 70 μs, the current amplitude is up to 40 kA. The septum knife thickness is 2.5 mm and it includes 0.5 mm ferromagnetic shield introduced to suppress leakage field in circulating beam vacuum chamber. The magnet cross-section is presented in Fig. 2.

---
[*] d.b.shwartz@inp.nsk.su

During machine commissioning phase the observation was done that circulating beams are disturbed by pulse of transfer channel components. Mainly this effect was due to closed loop of vacuum system and supports of additional ME4 magnet, similarly to what was observed in Ref. [14]. After the ceramic break was installed in transfer beamline's vacuum chamber to unclose loop the residual stray fields was related to septum magnet.

Many studies were done in different labs aiming the leakage field suppression [1-5, 15]. The accurate solution or simulation of this problem turned out to be quite difficult. The leakage field depends not only on the design of the septum and surroundings but also on the driving circuit. For example, VEPP-2000 septum is powered via lowering transformer. It means that secondary winding forms the closed loop after the driving pulse stopped with a switch in primary circuit.

It was realized that the main harmful effect of stray field was not the orbit shift but betatron tune sweep that pointed at list the inhomogeneous field. In addition, during booster BEP upgrade [16] the injection septum magnet was replaced by new one of similar to ME5 design. It's stray field was measured directly at the test bench and found to be significantly nonlinear (see Fig. 3).

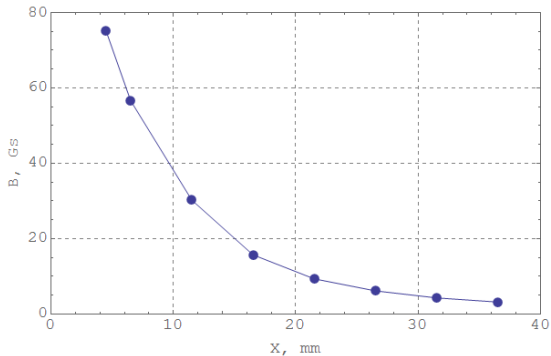

FIG. 3: Stray field of booster BEP pulsed injection septum magnet measured with small coil. Zero coordinate corresponds to septum knife surface.

Finally, it was proposed to make beam-based measurements of septa stray fields. The CO responses are routinely used at VEPP-2000 for the closed orbit measurement and correction, lattice tuning and final focus solenoids accurate alignment [17]. VEPP-2000 BPM system [18] is based on 16 CCD matrices reading beam profiles via synchrotron radiation (SR) outputs from each edge of dipoles (see Fig. 1). For pulsed CO distortions only the fast pickup BPMs are useful.

## III. TECHNIQUE

VEPP-2000 storage ring is equipped with only 4 pickups [19]. Each pickup works either in slow closed orbit measurement regime or in external trigger regime when electronics captures the set of turn-by-turn data. Maximal number of measured turns is 8192 but decimation can be applied to cover long-term period.

The typical pickup signal from the beam disturbed by septum stray field is shown in Fig. 4. The time of full-sine magnet driving pulse corresponds to 280 μs/ 82 ns = 3400 turns (shown with green zone in Fig. 4). One can see that stray field has much longer decay time.

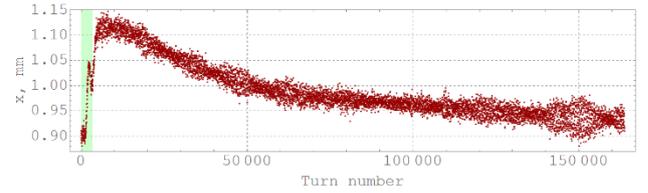

FIG. 4: Beam closed orbit distortion as seen by pickup with decimation (×20) switched on.

The single-turn beam position measurement accuracy does not exceed ~30 μm. Since we are interested in relatively slow orbit shift in order to increase the accuracy down to several microns 1000 turns are taken from pickup signal and averaged. In principle we can study the field distribution at any moment of decay, but we focused at the moment of maximal orbit distortion.

In order to extract stray filed value we first calculate the model response supposing that the leakage is well concentrated near thinnest part of the septum knife with length of ~10 cm. Both components of field $B_y$, $B_x$ can exist, thus responses in both planes should be anticipated (see Fig. 5). Although two septa for electron and positron injection are distanced less than 1 m the betatron phase advance is not small. The responses for ME5, MP5 differ significantly.

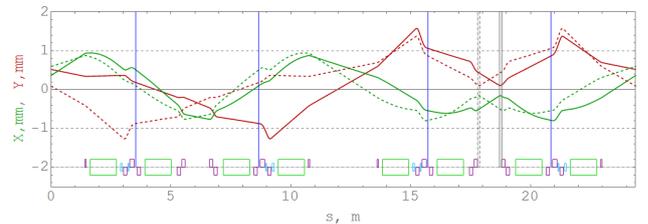

FIG. 5: Horizontal (green) and vertical (red) CO distortion caused by stray field of electron (solid lines) and positron (dashed lines) injection septa. Grey vertical solid- and dashed-edge stripes show the position of ME5, MP5 correspondingly. Blue lines indicate positions of pickups.

Response measured by four pickups $\vec{x}_{exp}$ or $\vec{y}_{exp}$ is compared with model vector of four components, and fitted amplitude gives the integral of disturbing field. The residual mean square difference allows to estimate the accuracy of field reconstruction. The example of fitted response is presented in Fig. 6.

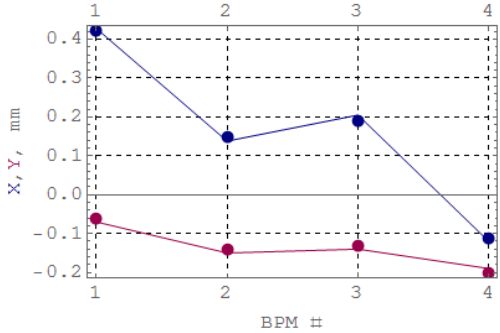

FIG. 6: Response fitting example. Points and lines are for measured and fitted CO shifts, correspondingly. Blue and purple are for horizontal and vertical responses.

Next step for the mapping is beam CO shift by DC correctors with precise control of the beam position at septum before the pulse. In our case due to small machine perimeter the CO distortion can't be done locally. In addition, to avoid the various chromatic effects in horizontal plane we prefer to use zero-integral field combination of steerers.

The static orbit distortion is controlled by CCD-based BPMs. Usually the single-beam regime is used for these experiments thus 8 BPMs are available for CO control. The pattern of CO distortion again is taken from the lattice model and compared to vector of 8 measured shifts. The mean square difference of experimental and fitted vectors used for estimation of reconstructed orbit shift error.

The precise coordinate reconstruction turned out to be the main difficulty. We do not use steerers calibration preferring the direct orbit control to avoid hysteresis issues. Nevertheless, for the large orbit distortions the nonlinearities go in action. First measurements were done in regular lattice with strong chromatic sextupoles and final focus solenoids which are known for nonlinear fringe fields. It was found that for large amplitudes orbit response pattern differs significantly from the model one. The model lattice first-order correction produced by known nonlinearities at distorted orbit only partially could explain this discrepancy.

Special "warm" lattice with switched off solenoids allowed us to improve the coordinate control precision. Here the overall focusing is relaxed, sextupoles are weaker and horizontal orbit distortion fortunately can be done almost zero at the chromatic sextupoles (see Fig. 7).

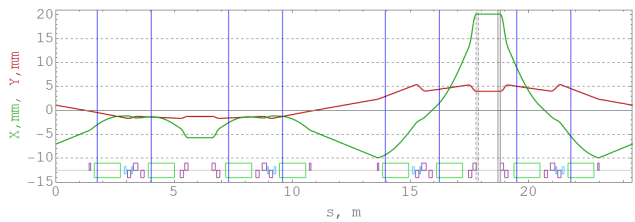

FIG. 7: Horizontal (green) and vertical (red) model CO pre-distortion used for coordinate scan.

We should mention that VEPP-2000 works at the main coupling resonance to keep the emittances equal. Even in "warm" regime the working point ($v_x$ = 2.44, $v_y$ = 1.44) was kept on the resonance. However, the orbit responses are not sensible to this resonance. This fact allow us to attribute horizontal orbit distortion to vertical magnetic field component only and vice versa.

Another obstacle for wide range coordinate scan is that beam image can escape from certain CCD's field of vision. One should vary the number of BPMs for fitting the CO shift value (see Fig. 8 for examples).

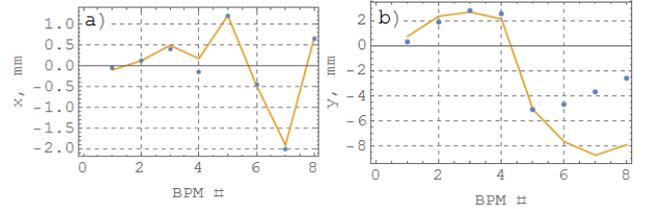

FIG. 8: CO pre-distortion fit examples: a) measured (points) and fitted (line) horizontal response of 8 BPMs; b) vertical response with three last BPMs lost the beam image.

It stands to reason that model-dependent approach of responses analysis demands the accurate machine tuning prior to measurements. Thus, firstly the orbit correction and lattice correction should be done.

## IV. RESULTS

The field mapping was carried out with circulating single electron beam at relatively low energy of 340 MeV. In Fig. 9-10 presented the measured maps of both field components with septum feeding pulse corresponding to regular injection at 340 MeV. Axis origin corresponds to design CO. The numbers show the field value in Gausses, with assumption that stray field region has length of 10 cm.

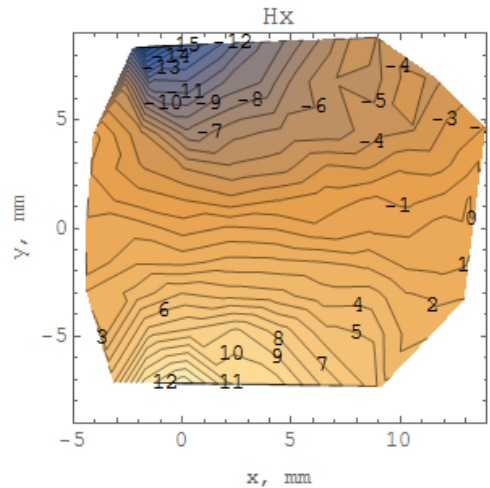

FIG. 9: Measured horizontal field component map.

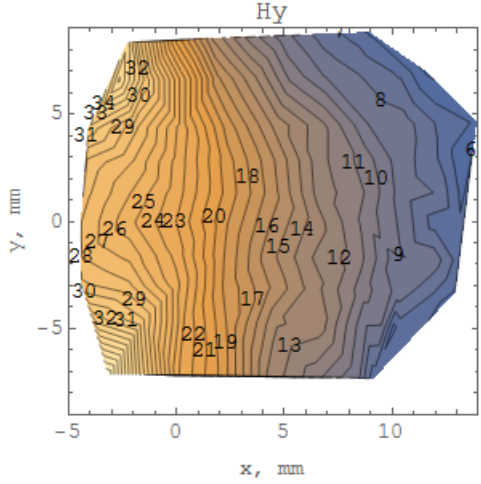

FIG. 10: Measured vertical field component map.

The estimated error of field fit is presented in Fig. 11a, it remains of order of 1 Gs. This small value indicates good agreement of experimental and model response. It supports the assumption of field leakage localized near septum knife and allows to throw away hypothesis of long-loop current transport along vacuum system. The coordinate error increase with orbit distortion up to 3 mm at the right border (see Fig. 11b).

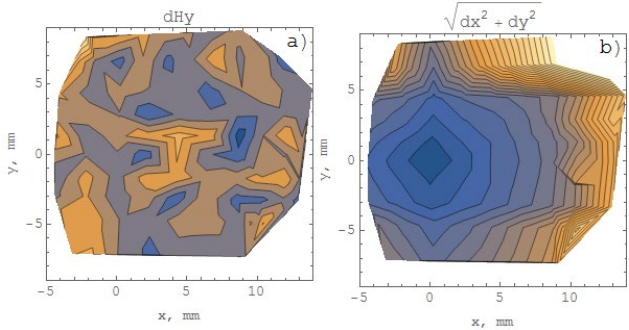

FIG. 11: Estimated errors of field (a) and coordinate (b) maps. Contours follow every 0.2 Gs and 0.2 mm correspondingly.

Representation form of measured field can be different, for example as a vector field (see Fig. 12).

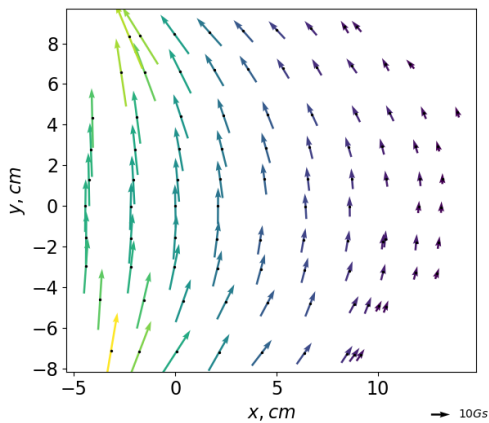

FIG. 12: Measured field shown as a vector field.

With given field maps one can numerically build the vector potential which has only one non-zero component $A_z$ since we assume 2D-field. The magnetic flux map can be figured as contour plot of $A_z$ as presented in Fig. 13.

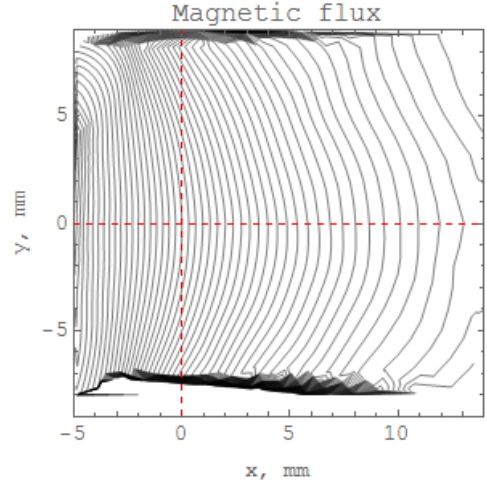

FIG. 13: Reconstructed magnetic flux. Red dashed cross shows the regular CO position.

In principle, the field being studied can have any 3D configuration. However, the beam CO response used for field detection is already proportional to field integral along the beam axis. The measured integrated 2D-field $\vec{B}_{exp}$ always can be presented as a superposition of 2D-multipoles [20,23]. In this case the field components along the circle with given radius R = 4 mm centred at the CO can be fitted as a series of multipole harmonics (see Fig. 14).

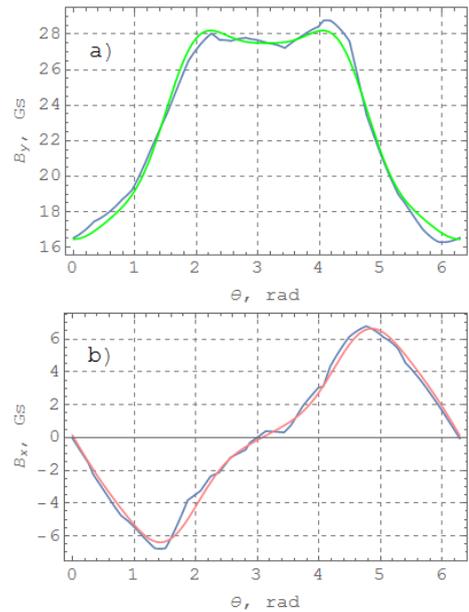

FIG. 14: Vertical (a) and horizontal (b) field components as a function of polar angle along the circle. Blue line for measured maps, green and red is a fitted sum of 5 multipoles.

In order to use maximum information of the measured data the full 2D-map was fitted for both field components simultaneously with a superposition of several multipoles up to octupole:

$$B_y + iB_x = \sum_{n=0}^{3} a_n \left( \text{Re}(x+iy)^n + i\,\text{Im}(x+iy)^n \right), \quad (1)$$

where $a_n$ are allowed to be complex numbers with imaginary part corresponding to skew multipoles. The fitted coefficients as can be predicted from the general appearance of previous figures showed that besides normal dipole and quadrupole components other multipoles are small. The resulting fitted field can be compared with measured maps in Fig. 15.

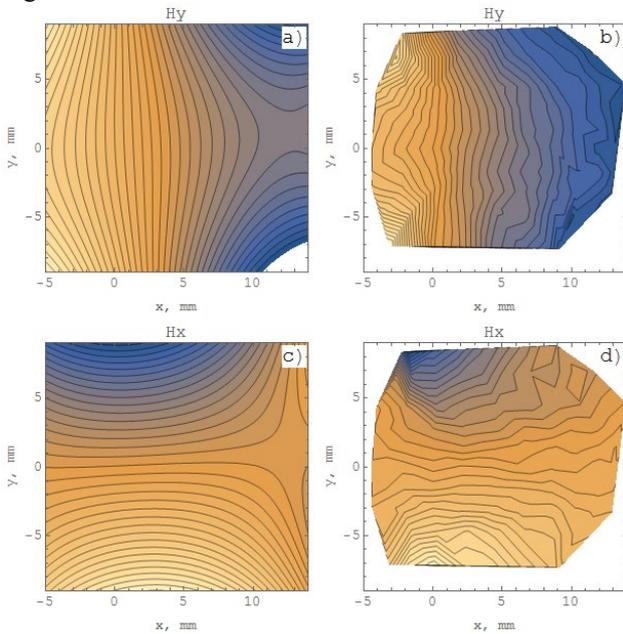

FIG. 15: Vertical (a, b) and horizontal (c, d) field components as measured (b, d) and fitted (a, c).

Another field map was measured for stronger septum magnet powering pulse corresponding to injected beam energy of 910 MeV (Fig. 16). The stray field is much higher and with approach close to the septum knife it killed the probing low energy 340 MeV beam. Thus, the horizontal coordinate is additionally limited from the left.

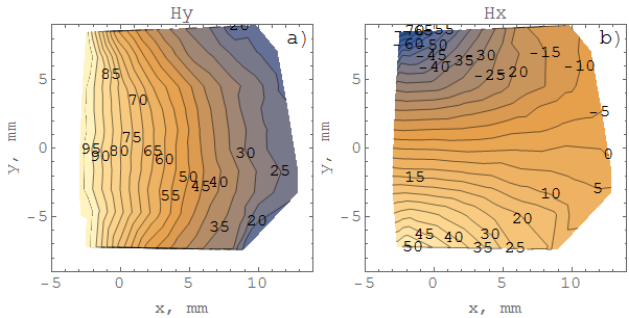

FIG. 16: Stray field map for the case of driving pulse amplitude close to maximum.

The dependence of stray field on the driving pulse amplitude at the unperturbed CO was measured (see Fig. 17). Although the dependence is nonlinear the significant slope at low current indicates that leakage is not explained by saturation of thin magnetic shield.

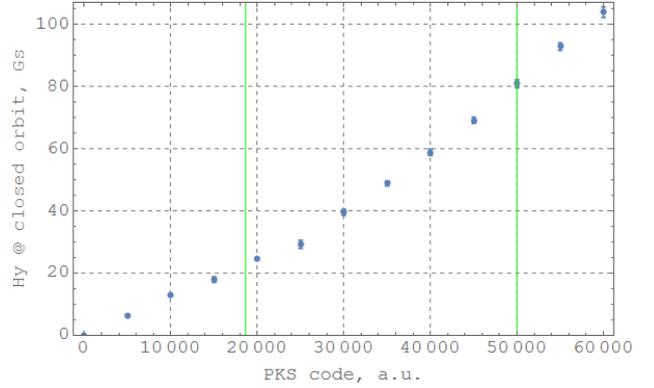

FIG. 17: Field at CO dependence on driving pulse amplitude. Vertical green lines marks the two levels used for full mapping.

Finally, the stray field of two different magnets ME5, MP5 was measured and compared. These injection septa for electron and positron beams correspondingly have identical design and both powered by single power converter with mechanical commutator. The dependences of stray field on horizontal coordinate for both magnets are shown in Fig. 18. It is clearly seen than positron septum for some reason has higher leakage field.

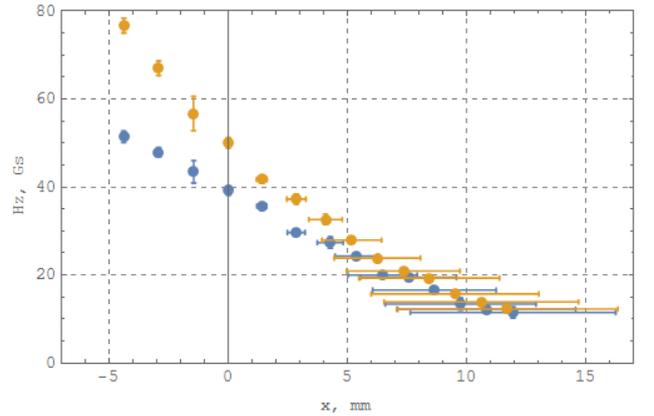

FIG. 18: Stray fields of ME5 (blue), MP5 (yellow) dependence on horizontal coordinate.

Let's remind that error bars in above figures indicate the estimation of systematic inaccuracy of the field and coordinate reconstruction. The statistical errors related to BPMs noise are small that is clear from error values in the vicinity of CO. In order to crosscheck the accuracy of large coordinate shift another approach was used. The electron beam lifetime at low energy is determined by Touschek scattering even for beam with low intensity of ~1 mA used for measurements. But for the limited transverse aperture, for example horizontal $A_x$, the elastic scattering on the nuclei of rest gas comes into play. For the very small aperture ~4-5 σ of beamsize the quantum lifetime gives the limit.

Thus, measurement of the lifetime $\tau$ with respect to orbit shift while approaching the known aperture limit can be used for coordinate calibration.

$$\frac{1}{\tau} = \frac{1}{\tau_{qu}} + \frac{1}{\tau_{el}}. \quad (2)$$

The lifetime driven by elastic scattering $\tau_{el}$ is proportional to square of aperture [22] thus can be expressed as:

$$\tau_{el} = \frac{A_x^2}{\alpha}, \quad (3)$$

where $\alpha$ coefficient depends on gas pressure and machine lattice functions. The quantum lifetime $\tau_{qu}$ has a very sharp dependence on the aperture limitation [23]:

$$\tau_{qu} = \tau_d \frac{\sigma_x^2}{A_x^2} e^{\frac{A_x^2}{2\sigma_x^2}}. \quad (4)$$

Here $\tau_d$ = 90 ms is damping time, $\sigma_x$ = 0.35 mm is horizontal beamsize at the aperture limitation region. Both this parameters are known from model. Beamsize additionally controlled at all CCDs to be in accordance with model. We will fit measured lifetime with a combination of two components with single aperture limit $x_0$.

$$\frac{1}{\tau} = \frac{1}{\tau_d} \frac{(x-x_0)^2}{\sigma_x^2} e^{-\frac{(x-x_0)^2}{2\sigma_x^2}} + \frac{\alpha}{(x-x_0)^2}. \quad (5)$$

In Fig. 19 the measured lifetime dependence is presented together with fit result.

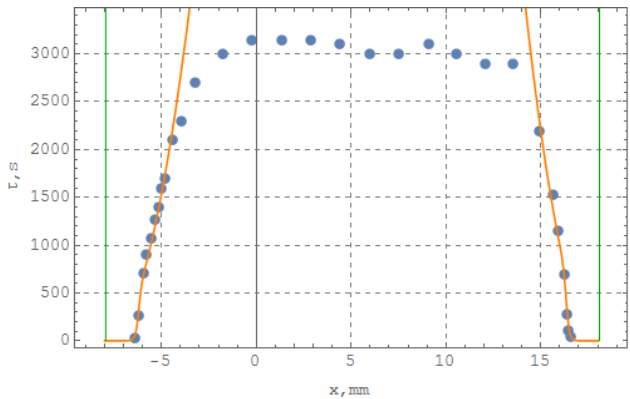

FIG. 19: Measured (points) and fitted (yellow lines) beam lifetime as a function of horizontal coordinate. Green lines indicate the position of obtained aperture limit.

The reconstructed from fit positions of mechanic aperture corresponds to $x_0 = -7.9$ mm, $x_0 = +18.1$ mm. Sum of these values gives the horizontal gap of 26 mm being in good agreement with design (see Fig. 2). This agreement indicates that coordinate errors were overestimated.

## V. CONCLUSION

In this paper we have shown that circulating beam can be used as a sensitive and convenient instrument for field mapping. Beam-based method allows to measure pulsed fields inside the vacuum chamber of assembled and operating machine. The stray field of VEPP-2000 septum magnets was mapped and analysed. As a rough fast cure a weak pulsed quadrupole corrector was constructed with pulse duration of ~10 ms to suppress the betatron tune excursion.

## ACKNOWLEDGEMENTS

We are grateful to D. E. Berkaev, B. Z. Persov, V. P. Prosvetov and Yu. M. Zharinov for support with pulsed septum design and operation. We would like to thank Yu. A. Rogovsky, A. L. Romanov and A. I. Senchenko for assistance in orbit responses technique software.

The work is partially supported by Russian Science Foundation under project N 14-50-00080.

## REFERENCES

[1] W. Kang, L. X. Fu, J. B. Pang, Development of an Eddy-Current Septum Magnet for the SSRF Storage Ring, *Proceedings of APAC'01*, Beijing, China (2001), p.663.

[2] D. Shuman *et al.*, Stray Field Reduction of ALS Eddy Current Septum Magnets, *Proceedings of PAC'05*, Knoxville, USA (2005), p.3718.

[3] P. Lebrasque *et al.*, Eddy Current Septum Magnets for Booster Injection and Extraction, and Storage Ring Injection at Synchrotron SOLEIL, *Proceedings of EPAC'06*, Edinburgh, Scotland (2006), p.3511.

[4] A. Zhuravlev *et al.*, Pulsed Magnets for Injection and Extraction Sections of NSLS-II 3 GeV Booster, *Proceedings of IPAC'13*, Shanghai, China (2013), p.3582.

[5] A. Ueda *et al.*, Construction of the New Septum Magnet Systems for PF-Advanced Ring, *Proceedings of IPAC'17*, Copenhagen, Denmark (2017), p.3398.

[6] B. Balhan *et al.*, Validation of the CERN PS Eddy Current Injection Septa, *Proceedings of IPAC'18*, Vancouver, Canada (2018), p.768.

[7] Yu. M. Shatunov *et al.*, Project of a New Electron-Positron Collider VEPP-2000, *Proceedings of EPAC'00*, Vienna, Austria (2000), p. 439.

[8] T. Mitsui *et al.*, Beam-Based Design of a Correction Coil for the Stray Field of a Pulse Septum Magnet", IEEE Transactions on Applied Superconductivity, Vol.18, Issue 2, pp. 1521-1524 (2008).

[9] D. E. Berkaev *et al.*, The VEPP-2000 electron-positron collider: First experiments, J. Exp. Theor. Phys., Vol.113, no.2, p.213 (2011).

[10] D. Shwartz *et al.*, Implementation of Round Colliding Beams Concept at VEPP-2000, *Proceedings of eeFACT'16*, Daresbury, UK (2016), p. 32.

[11] Yu. M. Shatunov *et al.*, Commissioning of the Electron-Positron Collider VEPP-2000 after the Upgrade, Phys. Part. Nucl. Lett. (2018) 15, p.310.

[12] V. V. Danilov *et al.*, The Concept of Round Colliding Beams, *Proceedings of EPAC'96*, Sitges, Spain (1996), p. 1149.

[13] D. E. Berkaev *et al.,* Beams Injection System for e$^+$e$^-$ Collider VEPP-2000, *Proceedings of EPAC'06*, Edinburgh, Scotland (2006), p.622.

[14] Y. Shoji, K. Kumagai, Unusual Eddy Current Stray Field of Pulse Septum of NewSUBARU, *Proceedings of The 14th Symposium on Accelerator Science and Technology,* Tsukuba, Japan (2003).

[15] K. Halbach, Some Thoughts on an Eddy Current Septum Magnet, Note LS-244 (1994).

[16] D. Shwartz *et al.*, Booster of Electrons and Positrons (BEP) Upgrade to 1 GeV, *Proceedings of IPAC'14*, Dresden, Germany (2014), p. 102.

[17] A. L. Romanov *et al.*, Round Beam Lattice Correction using Response Matrix at VEPP-2000, *Proceedings of IPAC'10*, Kyoto, Japan (2010), p. 4542.

[18] Yu. A. Rogovsky *et al.*, Beam Measurements with Visible Synchrotron Light at VEPP-2000 Collider, *Proceedings of DIPAC'11*, Hamburg, Germany (2011), p. 140.

[19] Yu. A. Rogovsky *et al.*, Pickup Beam Measurement System at the VEPP-2000 Collider, *Proceedings of RuPAC'10*, Protvino, Russia (2010), p. 101.

[20] K. Steffen, High energy beam optics (Interscience, New York, 1965).

[21] W. C. Elmore, M. W. Garret, Measurement of Two-Dimensional Fields, Rev. Sci. Instr., Vol.25, no.5, p.480 (1954).

[22] J. Le Duff, Current and Current Density Limitations in Electron Storage Rings, Nucl. Instrum. Methods Phys. Res., Sect. A 239, 83 (1985).

[23] A. W. Chao, M. Tigner, Handbook of Accelerator Physics and Engineering (WorldScientific, Singapore, 1999), p.188.